**Published in** ***Annals of Physics (NY),*** **262,** 47-72 (1998)

# Dirac's Representation Theory as a Framework for Signal Theory. I. Discrete Finite Signals

**Alexander Gersten**

**Department of Physics,**

**and Unit of Biomedical Engineering, and Zlotowski Center for Neuroscience**

**Ben-Gurion University of the Negev**

**Beer-Sheva 84105, Israel**

e-mail: gersten@bgumail.bgu.ac.il

**Abstract**

We demonstrate that the Dirac representation theory can be effectively adjusted and applied to signal theory. The main emphasis is on orthogonality as the principal physical requirement. The particular role of the identity and projection operators is stressed. A Dirac space is defined, which is spanned by an orthonormal basis labeled with the time points. An infinite number of orthonormal bases is found which are labeled with frequencies, they are distinguished by the continuous parameter $\alpha$. In a way, similar to one used in quantum mechanics, self-adjoint operators (observables) and averages (expectation values) are defined. Non orthonormal bases are discussed and it is shown, in an example, that they are less stable compared to the orthonormal ones. A variant of the sampling theorem for finite signals is derived. The aliasing phenomenon is described in the paper in terms of aliasing symmetry. Relations between different bases are derived. The uncertainty principle for finite signals is discussed.

## 1. Introduction

Signal theory is based largely on functional analysis and theory of distributions, on continuous and discrete systems. Quantum mechanics have similar mathematical problems [1], [2]. Moreover the distributions were employed by Dirac [1] in his representation theory before they gained a solid mathematical basis [3-5]. Dirac's representation theory is not just a different way of presenting known theories, it has its own special features which allow it to be extended to new theoretical problems [6] . Today it is almost impossible to think about new developments of quantum theory without the use of the transparent and deep formulations based on Dirac's representation theory. Such was, for example, Feynman's path integral formulation of quantum theories [7] . For a recent application see ref. [8]. The analogies of quantum theories and signal theory were pointed before [9],[10]. Here we concentrate mainly on the formulation of the basics of signal theory by using Dirac's representation theory. Elsewhere we discuss the feedback that signal theory may have on quantum theory. As the basics of signal theory require quite a large presentation, we were compelled to describe them in two papers, this paper, paper I, dealing with discrete finite signals, and paper II, dealing with infinite duration (continuous and discrete) signals [11].

In quantum mechanics, in one-dimension, the physical states corresponding to different eigenvalues are orthogonal. Therefore orthogonality is much more than a convenient tool for expansion of functions in terms of a convenient orthogonal basis. On the other hand the space of quantum states is a normed space and normalization of states is required. Usually, when the coordinates are extending to infinity, in order to secure normalizability, boundary conditions has to be imposed.
For scattering states, which are unbound, the situation is different, and one has to use generalized eigenfunctions, defined on rigged Hilbert spaces, which were initially developed by Gel'fand [12], [13], [14] , more elaborated by Maurin [15], and adapted to quantum mechanics by Roberts [16], Bohm [17], Antoine [18] and others [19].

The plan of our presentation is described below. We start in section 2 with the basics of Dirac's representation theory, trying to show its essentials by representing the bras and kets with rows and columns respectively. We introduce there the notion of coherent states and discuss the consequences of restricting the vector space to a subspace. In sec. 3 we introduce the "Dirac space" of dimension equal to the number of signal data. We discuss the orthogonal bases labeled with time and angular frequencies respectively. The analogy to the quantum wave functions is fully employed and averaged quantities are proposed. In sec. 4 we discuss nonorthogonal bases. In paper II we prove the well known sampling theorem using the methods of Dirac's representation theory by depicting it as a relation between the complete discrete time basis in the subspace, generated by restricting the frequencies, and the overcomplete continuous time basis. Here, in sec. 5, we give a different variant of the sampling theorem. We decompose the Dirac space into two subspaces of equal dimensions. The sampling theorem here is the relation between the complete basis in a subspace and the overcomplete basis obtained by projecting the basis of the Dirac space into the subspace. In sec. 6 we derive and discuss the symmetries which are leading to the aliasing phenomenon. In sec. 7 we introduce the signal and spectral operators respectively, and discuss their relation to the spectral amplitude and to double distributions. In sec. 8 some identities are derived. The uncertainty principle is the subject of sec. 9. In sec. 10 we summarize and discuss our findings.

## 2. A representation for the bra-kets

In Dirac's representation theory [1] the state (the system, the function or the object) are described by an abstract vector denoted as a ket vector $\left|p_1, p_2, \ldots\right\rangle$, where inside are the parameters which specify the state. Usually one can get results only after specifying the components of the ket vector. One can get them by projecting the vector into a set of unit vectors which form a basis for a linear complex space . In order to perform the projection we need to define a scalar product of two vectors.

This can be achieved by defining the dual space of bra vectors. Dual to the ket vector $|p_1, p_{2,}...\rangle$ is the bra vector $\langle p_1, p_2, ...|$ , which properties we describe below. In finite dimensions, the space of kets or bras is an abstract (separable) Hilbert space. In infinite dimensions one has to deal, for bound states, with the Hilbert space [2], and for unbound states, with rigged Hilbert spaces [12-19].

One can gain a better insight by considering finite dimensional vectors. For example

$$|a\rangle = \begin{pmatrix} a_1 \\ a_2 \\ \vdots \\ a_n \end{pmatrix}, \ \langle a| = \begin{pmatrix} a_1^* & a_2^* & \cdots & a_n^* \end{pmatrix} = |a\rangle^\dagger, \tag{2.1}$$

where $|a\rangle^\dagger$ is the hermitian conjugate (or the complex conjugate of the transpose) of $|a\rangle$. The scalar product can be defined in the usual way:

$$\langle a|a\rangle = \begin{pmatrix} a_1^* & a_2^* & \cdots & a_n^* \end{pmatrix} \begin{pmatrix} a_1 \\ a_2 \\ \vdots \\ a_n \end{pmatrix} = a_1^* a_1 + a_2^* a_2 + \cdots + a_n^* a_n \ . \tag{2.2}$$

An equivalent definition, in the spirit of Dirac's representation theory will be:

$$\begin{aligned} &\langle i|a\rangle = a_i, \ \ i = 1,2,\cdots n; \ \ \langle a|i\rangle = a_i^*; \ \ \langle i|j\rangle = \delta_{ij} = \begin{cases} 0 & \text{for } i \neq j \\ 1 & \text{for } i = j \end{cases} \\ &\langle a|a\rangle = \sum_i \langle a|i\rangle\langle i|a\rangle = a_1^* a_1 + a_2^* a_2 + \cdots + a_n^* a_n \end{aligned} \ . \tag{2.3}$$

Here the vectors $|i\rangle$ can be considered as a set of orthonormal vectors spanning an n dimensional basis for the vectors $|a\rangle$:

$$|a\rangle = a_1|1\rangle + a_2|2\rangle + \cdots + a_n|n\rangle \, . \tag{2.4}$$

The scalar product $\langle i|a\rangle$ is the projection of $|a\rangle$ in the direction $|i\rangle$.

The Dirac formalism allows, beside the scalar product, to consider the "dyadic" product of two vectors:

$$|a\rangle\langle b| = \begin{pmatrix} a_1 \\ a_2 \\ \vdots \\ a_n \end{pmatrix} \begin{pmatrix} b_1^* & b_2^* & \cdots & b_n^* \end{pmatrix} = \begin{pmatrix} a_1 b_1^* & a_1 b_2^* & \cdots & a_1 b_n^* \\ a_2 b_1^* & a_2 b_2^* & \cdots & a_2 b_n^* \\ \vdots & \vdots & \ddots & \\ a_n b_1^* & a_n b_2^* & \cdots & a_n b_n^* \end{pmatrix} . \tag{2.5}$$

From eqs. (2.5) and (2.3), for an orthonormal basis, one can get:

$$I \equiv \sum_{k=1}^{n} |k\rangle\langle k| = \begin{pmatrix} 1 & & & \\ & 1 & & \\ & & \ddots & \\ & & & 1 \end{pmatrix}, \quad (n \times n \text{ unit matrix}). \tag{2.6}$$

One can consider the expression in eq. (2.6) as the definition of the identity operator I, and rewrite eqs. (2.1) and (2.2) in the form:

$$\langle a|a\rangle = \langle a||I|a\rangle = \langle a|( \sum_k |k\rangle\langle k| )|a\rangle = a_1^* a_1 + a_2^* a_2 + \cdots + a_n^* a_n . \tag{2.7}$$

Equation (2.6) represents the completeness relation, in the sense, that any vector in the n dimensional ket space can be expanded uniquely with the basis $|k\rangle$:

$$|a\rangle = I|a\rangle = \left( \sum_{k=1}^{n} |k\rangle\langle k| \right)|a\rangle = \sum_{k=1}^{n} \langle k|a\rangle|k\rangle \equiv \sum_{i=1}^{n} a_k |k\rangle . \tag{2.8}$$

It may be advantageous to introduce the operator (a matrix):

$$F = \sum_{k=1}^{n} |k\rangle f_k \langle k| , \tag{2.9}$$

which we will call the "filter operator". It can reduce the dimensions of the considered space if some of the $f_i$ are equal to zero. We shall apply such operators in signal analysis when constraints will be put on the frequency space basis. A particular case of a filter operator is the projection operator P which satisfies the condition:

$$P = P^{\dagger} = P^2 . \tag{2.10}$$

In a way similar to the definition of the bra as the hermitian conjugate of the ket in eq. (2.1), one can define the adjoint of an operator (matrix). If A is an operator whose action is given by

$$A|a\rangle = |b\rangle , \tag{2.11}$$

then the adjoint operator is defined by:

$$\langle b| = \langle a|A^{\dagger}. \tag{2.12}$$

Other definition of the adjoint operator follows the relation:

$$\langle b|A|a\rangle = \langle a|A^{\dagger}|b\rangle. \tag{2.13}$$

Further simplifications are possible if we note that for a projection operator

$$P = P^{\dagger}. \tag{2.14}$$

Eq. (2.14) can be proved easily if we note that from eq. (2.10) by conjugation we obtain:

$$P^{\dagger} = (P^{\dagger})^2. \tag{2.15}$$

Combining eqs. (2.10) and (2.15) we obtain:

$$P^2P^{\dagger}=P(P^{\dagger})^2,$$

which reduces to eq. (2.14). An important example of a projection operator for the orthonormal basis (2.6) is

$$P = \sum_{k=1}^{n} |k\rangle p_k \langle k|, \tag{2.16}$$

where the coefficients $p_k$ are either equal to 1 or to 0. Let us consider two examples.

**Example 1.** Consider two orthonormal bases in a three dimensional space $C^3$ (of complex numbers):

$$|1\rangle = \begin{pmatrix}1\\0\\0\end{pmatrix}, \quad |2\rangle = \begin{pmatrix}0\\1\\0\end{pmatrix}, \quad |3\rangle = \begin{pmatrix}0\\0\\1\end{pmatrix}, \quad \text{and:}$$
$$|1'\rangle = \frac{1}{\sqrt{2}}\begin{pmatrix}1\\0\\i\end{pmatrix}, \quad |2'\rangle = \begin{pmatrix}0\\1\\0\end{pmatrix}, \quad |3'\rangle = \frac{1}{\sqrt{2}}\begin{pmatrix}1\\0\\-i\end{pmatrix}, \tag{2.17}$$

with the identity operator (unit matrix):

$$I = \sum_{k=1}^{3} |k\rangle\langle k| = \sum_{k=1}^{3} |k'\rangle\langle k'| = \begin{pmatrix}1&0&0\\0&1&0\\0&0&1\end{pmatrix}. \tag{2.18}$$

Let us impose now a constraint that all vectors of the system are confined to a subspace $C_F^2$ spanned by the basis vectors $|1'\rangle$ and $|2'\rangle$. This constraint will be realized with the help of the filter operator (now also a projection operator):

$$F = |1'\rangle\langle 1'| + |2'\rangle\langle 2'| = F^{\dagger}, \tag{2.19}$$

where $F^{\dagger}$ is the adjoint operator (or hermitian conjugate operator).

Any three dimensional vector $|a\rangle$

$$|a\rangle = a_1|1\rangle + a_2|2\rangle + a_3|3\rangle = a_{1'}|1'\rangle + a_{2'}|2'\rangle + a_{3'}|3'\rangle, \tag{2.20}$$

will be projected into the required subspace $C_F^2$ by the operation (eq. 2.19):

$$\begin{aligned} F|a\rangle &= a_{1'}|1'\rangle + a_{2'}|2'\rangle = a_1F|1\rangle + a_2F|2\rangle + a_3F|3\rangle \\ &= a_1(\tfrac{1}{\sqrt{2}}|1'\rangle) + a_2(|2'\rangle) + a_3(-\tfrac{i}{\sqrt{2}}|1'\rangle). \end{aligned} \tag{2.21}$$

One can use eq. (2.18) and note here an interesting result for the identity operator $I_F$ in the constrained subspace $C_F^2$ :

$$I_F = |1'\rangle\langle 1'| + |2'\rangle\langle 2'| = FI\ F^{\dagger} = F|1\rangle\langle 1|\ F^{\dagger} + F|2\rangle\langle 2|F^{\dagger} + F|3\rangle\langle 3|F^{\dagger}, \tag{2.22}$$

i.e. the states $F|k\rangle$, (k=1,2,3), form an overcomplete nonorthogonal basis for $C_F^2$ .

Any vector in $C_F^2$, can be expanded in a unique way in this basis:

$$|b\rangle = b_{1'}|1'\rangle + b_{2'}|2'\rangle = b_1F|1\rangle + b_2F|2\rangle + b_3F|3\rangle, \tag{2.23}$$

and from eq. (2.22) we obtain:

$$|b\rangle = I_F|b\rangle = \langle 1|F^{\dagger}|b\rangle F|1\rangle + \langle 2|F^{\dagger}|b\rangle F|2\rangle + \langle 3|F^{\dagger}|b\rangle F|3\rangle, \tag{2.24}$$

with the expansion coefficients

$$b_k = \langle k|F^{\dagger}|b\rangle. \tag{2.25}$$

Comparing eqs. (2.23) with eq. (2.21) we can see that a vector in $C^3$ will have the same expansion coefficients in $C_F^2$ if the (overcomplete nonorthogonal) basis given by eq. (2.24) will be used. States which form an overcomplete nonorthogonal basis with a continuous basis index, are being referred in ref. [20] as "coherent states". Here we shall use this term also for a discrete basis.

Now we are able to generalize the previous example and formulate the following theorem:

**Example 2. The Projection Theorem.**

Let $|k\rangle$ be an orthonormal n dimensional basis in $C^n$, satisfying the completeness relation $I_n = \sum_{k=1}^{n} |k\rangle\langle k|$, where $I_n$ is the identity operator in $C^n$. Let $C^m$ be an m dimensional subspace of $C^n$ (m<n) spanned by the orthonormal basis $|k'\rangle$ and the corresponding identity operator

$$I_m = \sum_{k'=1}^{m} |k'\rangle\langle k'| . \quad (2.26)$$

Let P be a projection operator projecting vector states from $C^n$ into vector states of $C^m$. Then any vector state $|b\rangle_m$ of $C^m$ can be expanded in $C^n$ according to

$$|b\rangle_n = \sum_{k=1}^{n} b_k |k\rangle , \quad (2.27)$$

where

$$b_k = \langle k|P^{\dagger}|b\rangle_m = \sum_{k'=1}^{m} \langle k|P^{\dagger}|k'\rangle\langle k'|b\rangle_m , \qquad k = 1,2,\ldots n , \quad (2.28)$$

and

$$I_m = \sum_{k=1}^{n} P|k\rangle\langle k|P^{\dagger} , \quad (2.29)$$

i.e. the expansion coefficients $b_k$ are calculated in $C^m$ using the coherent basis $P|k\rangle$. The second part of eq. (2.28) was obtained by using eq. (2.26).

## 3. Orthonormal bases in the Dirac space

In practice, the analog signal u(t) is being usually sampled, every time interval Δt, with a sampling rate:

$$f_S = 1/\Delta t . \quad (3.1)$$

Let us assume that the sampling contains K data points at times:

$$t_k = k\,\Delta t, \; k=1,2,\ldots, K . \quad (3.2)$$

Following the ideas of Dirac [1] we can represent the sampled signal $u(t_k)$ (k=1,2,...K) by the ket-vector in the K dimensional vector space:

$$|u\rangle = v\begin{pmatrix} u(t_1) \\ u(t_2) \\ \vdots \\ u(t_K) \end{pmatrix}, \quad (3.3)$$

which we will call the Dirac space. Above $\nu$ is a normalization constant to be discussed later on. The Dirac space is spanned by the canonical basis:

$$|t_1\rangle = \begin{pmatrix}1\\0\\\vdots\\0\end{pmatrix}, \quad |t_2\rangle = \begin{pmatrix}0\\1\\\vdots\\0\end{pmatrix}, \quad \cdots \quad |t_K\rangle = \begin{pmatrix}0\\0\\\vdots\\1\end{pmatrix}. \tag{3.4}$$

so that

$$\langle t_k | u\rangle = \nu u(t_k). \tag{3.5}$$

If we normalize the ket (3.3) to unity, then

$$\nu = 1 \Big/ \sqrt{\sum_{k=1}^{K} u^*(t_k) u(t_k)}, \tag{3.5a}$$

and the ket (3.3) will have a similar significance as wavefunction in quantum mechanics and we will be able to speak about probability amplitudes. The scalar product:

$$\langle u | u\rangle = \sum_{k=}^{K} \left[\nu^* u^*(t_k)\right]\left[\nu u(t_k)\right] = 1, \tag{3.5b}$$

has the meaning that the sum of all probabilities must equal one. The identity operator (the unit matrix) can be written as:

$$I = \sum_{k=1}^{K} |t_k\rangle\langle t_k|. \tag{3.6}$$

In order to perform spectral analysis we introduce a new basis for the Dirac space

$$|\omega_k\rangle, \ k=1, 2,\ldots,K, \tag{3.7}$$

linked with the angular frequency projections. The basis (3.7) is connected with the basis (3.4) via the matrix:

$$\langle t_n | \omega_k\rangle = \eta \exp(i\omega_k t_n), \tag{3.8}$$

where the angular frequencies $\omega_k$ will be determined later in accord with the discrete Fourier transform (DFT). Thus in the Dirac space this new basis will have the following components:

$$\left|\omega_k\right\rangle = I\left|\omega_k\right\rangle = \sum_{n=1}^{K}\left\langle t_n \middle| \omega_k\right\rangle\left|t_n\right\rangle = \eta\begin{pmatrix}\exp(i\omega_k t_1)\\ \exp(i\omega_k t_2)\\ \vdots \\ \exp(i\omega_k t_K)\end{pmatrix}. \tag{3.9}$$

We will require orthonormality of this basis:

$$\left\langle \omega_n \middle| \omega_m \right\rangle = \delta_{nm}, \tag{3.10}$$

From eqs. (3.2), (3.6) and (3.9) we get:

$$\left\langle \omega_n \middle| \omega_m \right\rangle = \sum_{k=1}^{K}\left\langle \omega_n \middle| t_k\right\rangle\left\langle t_k \middle| \omega_m\right\rangle = \left|\eta\right|^2 \sum_{k=1}^{K} e^{i(\omega_m-\omega_n)k\Delta t}. \tag{3.11}$$

The sum can be evaluated using the summation formulae of the geometric series:

$$1 + x + x^2 + \ldots + x^{K-1} = \frac{x^K - 1}{x-1}, \tag{3.12.a}$$

$$x^{\alpha} + x^{\alpha+1} + x^{\alpha+2} + \ldots + x^{\alpha+K-1} = x^{\alpha}\frac{x^K - 1}{x-1}, \tag{3.12b}$$

where eq. (3.12.b) was obtained by multiplying eq. (3.12b) by $x^{\alpha}$. We obtain:

$$\left\langle \omega_n \middle| \omega_m \right\rangle = \left|\eta\right|^2 e^{i(\omega_m-\omega_n)\Delta t}\frac{e^{i(\omega_m-\omega_n)K\Delta t} - 1}{e^{i(\omega_m-\omega_n)\Delta t} - 1}, \tag{3.13}$$

and the orthonormality conditions will be:

$$\begin{aligned}\omega_n - \omega_m &= \frac{2\pi(n-m)}{K\Delta t} = (n-m)\Delta\omega,\\ \Delta\omega &= 2\pi(K\Delta t)^{-1} = 2\pi f_S / K,\\ \eta &= \frac{1}{\sqrt{K}}.\end{aligned} \tag{3.14}$$

In eq. (3.14) the difference n-m is an integer, but n and m separately do not have to be integers, one can add to them an overall real constant, denoted below as $\alpha$, which will be canceled when differences will be taken. Therefore we may choose:

$$\omega_{n,\alpha} = (\alpha + n - 1)\Delta\omega, \quad n = 1,2,\ldots,K, \tag{3.15a}$$

where $\alpha$ is the overall constant. The identity operator (the unit matrix) will take the form:

$$I=\sum_{k=1}^{K} \left|\omega_{k,\alpha}\right\rangle\left\langle\omega_{k,\alpha}\right|, \tag{3.15b}$$

i.e. there are infinite number of orthonormal bases, characterized by the real constant $\alpha$, with which one can cover any angular frequency. The choice $\alpha=0$ is consistent with the standard DFT (except for normalization, $\eta=1/K$ for DFT) for which we have an orthonormal basis for the angular frequencies:

$$\left|\omega_k\right\rangle=\frac{1}{\sqrt{K}}\begin{pmatrix}\exp(i\omega_k t_1)\\ \exp(i\omega_k t_2)\\ \vdots\\ \exp(i\omega_k t_K)\end{pmatrix}, \qquad \omega_k=(k-1)\Delta\omega, \quad k=0,1,2,\ldots K-1. \tag{3.16}$$

and the identity operator (the unit matrix) will take the form:

$$I=\sum_{k=1}^{K} \left|\omega_k\right\rangle\left\langle\omega_k\right|. \tag{3.17}$$

The spectral amplitude, for this case, will be:

$$U(\omega_k)\equiv\left\langle\omega_k\middle|u\right\rangle=\sum_{n=1}^{K}\left\langle\omega_k\middle|t_n\right\rangle\left\langle t_n\middle|u\right\rangle=\frac{1}{\sqrt{K}}\sum_{n=1}^{K}\nu u(t_n)\exp(-i\omega_k t_n). \tag{3.18}$$

For $\alpha$ different from zero instead of eq. (3.16) we have:

$$\left|\omega_{k,\alpha}\right\rangle=\frac{1}{\sqrt{K}}\begin{pmatrix}\exp(i(\omega_k+\alpha\Delta\omega)t_1)\\ \exp(i(\omega_k+\alpha\Delta\omega)t_2)\\ \vdots\\ \exp(i(\omega_k+\alpha\Delta\omega)t_K)\end{pmatrix}, \tag{3.16a}$$

and more generally:

$$U(\omega_{k,\alpha})\equiv\left\langle\omega_{k,\alpha}\middle|u\right\rangle=\sum_{n=1}^{K}\left\langle\omega_{k,\alpha}\middle|t_n\right\rangle\left\langle t_n\middle|u\right\rangle=\frac{\nu}{\sqrt{K}}\sum_{n=1}^{K}u(t_n)\exp(-i\omega_{k,\alpha}t_n), \tag{3.18a}$$

and its inverse, the signal, with its spectral decomposition

$$u(t_k)\equiv\frac{\left\langle t_k\middle|u\right\rangle}{\nu}=\frac{1}{\nu}\sum_{n=1}^{K}\left\langle t_k\middle|\omega_{n,\alpha}\right\rangle\left\langle\omega_{n,\alpha}\middle|u\right\rangle=\frac{1}{\nu\sqrt{K}}\sum_{n=1}^{K}U(\omega_{n,\alpha})\exp(i\omega_n t_k), \tag{3.19}$$

where, for the derivation of eqs. (3.18-9), eqs. (3.6), (3.15b) and (3.17) were used. It is interesting to note that by choosing different values of $\alpha$ one can reach all values of the angular frequency $-\infty\le\omega\le\infty$, therefore eq. (3.18) can be used as both an interpolating and an extrapolating formula for the principal frequencies chosen with a specific value of α (like eq. (3.16)). In this case we can rewrite eq. (3.18a) in the following way:

$$U(\omega) \cong \langle \omega | u \rangle = \sum_{n=1}^{K} \langle \omega | t_n \rangle \langle t_n | u \rangle = \frac{\nu}{\sqrt{K}} \sum_{n=1}^{K} u(t_n) \exp(-i\omega t_n), \tag{3.20}$$

which, for real signals, has the following symmetry:

$$U^*(\omega) = U(-\omega). \tag{3.21}$$

The spectral power is defined as:

$$P(\omega) = |U(\omega)|^2. \tag{3.22}$$

The spectral power, for real signals, can also be evaluated as follows:

$$\begin{aligned} P(\omega) &= \langle u|\omega\rangle\langle\omega|u\rangle = \sum_{n=1}^{K} \langle u|t_n\rangle\langle t_n|\omega\rangle \sum_{k=1}^{K} \langle\omega|t_k\rangle\langle t_k|u\rangle \\ &= \frac{\nu^2}{K}\sum_{n=1}^{K}\sum_{k=1}^{K} u(t_n)u(t_k)e^{i\omega(t_n-t_k)} = \frac{\nu^2}{K}\sum_{n=1}^{K}\sum_{k=1}^{K} u(t_n)u(t_k)\cos(t_n - t_k). \end{aligned} \tag{3.23}$$

An example for the power dependence on frequency, using eqs. (3.22) or (3.23), is presented in fig. 1, where the signal was generated by:

$$u(t_k) = \cos(2\pi f_1 t_k + .3) + 0.6 \cdot \cos(2\pi f_2 t_k + .5) + 1.4 \cdot \cos(2\pi f_3 t_k + .8), \tag{3.24}$$

with: $f_1$=5 Hz, $f_2$=5.4 Hz, $f_3$=44 Hz, k=1,2,…50, $f_S$=30 Hz=1/(Δt), $t_k$=kΔ .

The appearing symmetries (aliasing) will be explained in sec. 7. Using eqs. (3.5a), (3.5b), (3.15b), (3.18a) we obtain the Parseval theorem in the following form:

$$\begin{aligned} \langle u|u\rangle &= \sum_{k=1}^{K} \left[\nu^* u^*(t_k)\right]\left[\nu u(t_k)\right] = \sum_{k=1}^{K} \langle u|(\omega_{k,\alpha})\rangle\langle(\omega_{k,\alpha})|u\rangle \\ &= \sum_{k=1}^{K} U^*(\omega_{k,\alpha})U(\omega_{k,\alpha}) = 1, \end{aligned} \tag{3.25}$$

with a total power normalized to one. From eqs. (3.18a) and (3.25) we can give $U(\omega_{k,\alpha})$ the meaning of an angular frequency probability amplitude. With this definition experiments with a different K (number of data points) will have similar values, if the same quantity is measured (unlike the standard DFT). Likewise using probability amplitudes allows to define more simply and profoundly various average quantities. For example the average angular frequency for the signal can be defined as $\langle u|\hat{\omega}_\alpha|u\rangle / \langle u|u\rangle$, where $\hat{\omega}_\alpha$ is the angular frequency operator, which will be defined in the section 4. The situation is somewhat complicated due to the fact that there are infinitely many orthonormal bases (3.15b) characterized by the parameter $\alpha$ via eq. (3.15a). Thus:

$$\langle \hat{\omega}_\alpha \rangle = \frac{\langle u|\hat{\omega}_\alpha|u\rangle}{\langle u|u\rangle}, \tag{3.26}$$

is the average angular frequency in the range:

$$\omega_{1,\alpha} \le \omega \le \omega_{K,\alpha} . \tag{3.27}$$

Eq. (3.21) is the symmetry of the Fourier transform of real signals. Therefore the basis, which from the point of view of symmetries, is the closest to the Fourier transform, seems to be the one with:

$$\alpha = -(K-1)/2 \equiv s \tag{3.28}$$

$$\omega_{n,s} = (n - K/2 - 1/2)\Delta\omega, \;\; n = 1,2,\ldots,K \; , \tag{3.29}$$

which for real signals has the same symmetry as eq. (3.21), namely:

$$U^*(\omega_{k,s}) = U(-\omega_{k,s}) = U(\omega_{K-k+1,s}), \;\; k = 1,2,\ldots,K, \tag{3.30}$$

and therefore only half of the amplitudes has to be computed.

## 4. Operators, averages and double distributions

In a similar way as observables are defined in quantum mechanics we can asign a self-adjoint operator, say $\hat{A}$, to a measured signal quantity and evaluate its average value according to:

$$\langle \hat{A} \rangle = \frac{\langle u|\hat{A}|u\rangle}{\langle u|u\rangle} . \tag{4.1}$$

Let us give few important examples:

**Example 1**: The time operator $\hat{t}$ will be defined as:

$$\hat{t} = \sum_{k=1}^{K} t_k |t_k\rangle\langle t_k| . \tag{4.2}$$

It has the following obvious properties:

$$\hat{t}|t_k\rangle = t_k|t_k\rangle, \tag{4.3}$$

$$\hat{t}^n = \sum_{k=1}^{K} (t_k)^n |t_k\rangle\langle t_k|. \tag{4.4}$$

The average time will be defined as:

$$\langle\hat{t}\rangle = \langle u|\hat{t}|u\rangle / \langle u|u\rangle = \sum_{k=1}^{K} t_k \langle u|t_k\rangle\langle t_k|u\rangle / \langle u|u\rangle = \sum_{k=1}^{K} t_k v^2 |u(t_k)|^2. \tag{4.5}$$

**Example 2**: the angular frequency operator can be defined as:

$$\hat{\omega}_\alpha = \sum_{k=1}^{K} \omega_{k,\alpha}|\omega_{k,\alpha}\rangle\langle\omega_{k,\alpha}|, \tag{4.6}$$

where $\alpha$ is a continuous parameter labeling the different orthogonal bases as in eq. (3.15a). The operator has the obvious properties:

$$\hat{\omega}_\alpha|\omega_{k,\alpha}\rangle = \omega_{k,\alpha}|\omega_{k,\alpha}\rangle, \tag{4.7}$$

$$(\hat{\omega}_\alpha)^n = \sum_{k=1}^{K} (\omega_{k,\alpha})^n |\omega_{k,\alpha}\rangle\langle\omega_{k,\alpha}|. \tag{4.8}$$

The situation is complicated due to the fact that there are infinitely many orthonormal bases (3.15b) characterized by the parameter $\alpha$.

The operator has a simple form in the $\omega$-basis. In the t-basis it will aquire the following form:

$$\begin{aligned}\hat{\omega}_\alpha &= \sum_{k=1}^{K} \omega_{k,\alpha}|\omega_{k,\alpha}\rangle\langle\omega_{k,\alpha}| = \sum_{m-1}^{K}\sum_{n=1}^{K}\sum_{k=1}^{K} \omega_{k,\alpha}\langle\omega_{k,\alpha}|t_n\rangle\langle t_m|\omega_{k,\alpha}\rangle|t_n\rangle\langle t_m| \\ &= \frac{1}{K}\sum_{m-1}^{K}\sum_{n=1}^{K}\sum_{k=1}^{K} \omega_{k,\alpha} e^{i\omega_{k,\alpha}(t_m - t_n)}|t_n\rangle\langle t_m|.\end{aligned} \tag{4.9}$$

Its average value is equal to:

$$\begin{aligned}\langle\hat{\omega}_\alpha\rangle &= \frac{\langle u|\hat{\omega}_\alpha|u\rangle}{\langle u|u\rangle} = \sum_{k=1}^{K} \frac{\langle u|\omega_{k,\alpha}\rangle\omega_{k,\alpha}\langle\omega_{k,\alpha}|u\rangle}{\langle u|u\rangle} \\ &= \sum_{k=1}^{K} \frac{U^*(\omega_{k,\alpha})\omega_{k,\alpha}U(\omega_{k,\alpha})}{\langle u|u\rangle},\end{aligned} \tag{4.10}$$

where the angular frequency is in the range:

$$\omega_{1,\alpha} \le \omega \le \omega_{K,\alpha}. \tag{4.11}$$

**Example 3**: Let us consider the following operator:

$$\hat{u} = \sum_{n=1}^{K} u(t_n)|t_n\rangle\langle t_n|, \tag{4.12}$$

the "signal operator", which has the following properties:

$$\hat{u}|t_m\rangle = u(t_m)|t_m\rangle, \tag{4.13}$$

$$\hat{u}\hat{u}^{\dagger} = \sum_{n=1}^{K}\sum_{k=1}^{K} u(t_n)u^*(t_k)|t_n\rangle\langle t_n|t_k\rangle\langle t_k| = \sum_{n=1}^{K} |u(t_n)|^2|t_n\rangle\langle t_n|, \tag{4.14}$$

$$\mathrm{tr}(\hat{u}\hat{u}^{\dagger}) = \sum_{n=1}^{K} |u(t_n)|^2, \tag{4.15}$$

Let us consider the transition matrix element:

$$\begin{aligned}\langle\omega|\hat{u}|0\rangle &= \sum_{n=1}^{K} \langle\omega|t_n\rangle\langle t_n|0\rangle u(t_n) = \frac{1}{K}\sum_{n=1}^{K} e^{-i\omega t_n}u(t_n)\\ &= \frac{1}{\sqrt{K}}\langle\omega|u\rangle = \frac{1}{\sqrt{K}}U(\omega).\end{aligned} \tag{4.16}$$

Thus the transition matrix of the signal operator between zero and ω angular frequencies is proportional to the spectral amplitude.

**Example 4**: We can define also the "spectral operator":

$$\hat{U}_{\alpha} = \sum_{n=1}^{K} U(\omega_{n,\alpha})|\omega_{n,\alpha}\rangle\langle\omega_{n,\alpha}|, \tag{4.17}$$

which satisfies:

$$\hat{U}_{\alpha}|\omega_{n,\alpha}\rangle = U(\omega_{n,\alpha})|\omega_{n,\alpha}\rangle, \tag{4.18}$$

$$\hat{U}_{\alpha}\hat{U}^{\dagger}_{\alpha} = \sum_{n=1}^{K} |U(\omega_{n,\alpha})|^2|\omega_{n,\alpha}\rangle\langle\omega_{n,\alpha}|, \tag{4.19}$$

$$\mathrm{tr}(\hat{U}_{\alpha}\hat{U}^{\dagger}_{\alpha}) = \sum_{k=1}^{K}\sum_{n=1}^{K} |U(\omega_{n,\alpha})|^2\langle t_k|\omega_{n,\alpha}\rangle\langle\omega_{n,\alpha}|t_k\rangle = \frac{1}{K}\sum_{n=1}^{K} |U(\omega_{n,\alpha})|^2. \tag{4.20}$$

Let us consider the transition matrix element:

$$\langle t_m|\hat{U}_{\alpha}|0\rangle = \sum_{n=1}^{K} \langle t_m|\omega_{n,\alpha}\rangle\langle\omega_{n,\alpha}|0\rangle U(\omega_{n,\alpha}) = \frac{1}{\sqrt{K}}\langle t_m|u\rangle = \frac{1}{\sqrt{K}}u(t_m), \tag{4.21}$$

with a result symmetric to eq. (4.16).

**Example 5**: One can combine the operators of example 3 and example 4:

$$\hat{U}_\alpha \hat{U}^\dagger{}_\alpha \hat{u}\hat{u}^\dagger = \sum_{k=1}^{K} \sum_{n=1}^{K} \left|U(\omega_{n,\alpha})\right|^2 \left|u(t_k)\right|^2 \left|\omega_{n,\alpha}\right\rangle\left\langle\omega_{n,\alpha}\right|t_k\rangle\langle t_k|, \tag{4.22}$$

$$\left\langle\omega_{m,\alpha}\right| \hat{U}_\alpha \hat{U}^\dagger{}_\alpha \hat{u}\hat{u}^\dagger \left|t_n\right\rangle = \left|U(\omega_{m,\alpha})\right|^2 \left|u(t_n)\right|^2 \left\langle\omega_{m,\alpha}\middle|t_n\right\rangle. \tag{4.23}$$

Equation (4.23) may form a basis for a double distribution:

$$W(t_n, \omega_{m,\alpha}) = \left\langle\omega_{m,\alpha}\right| \hat{U}_\alpha \hat{U}^\dagger{}_\alpha \hat{u}\hat{u}^\dagger \left|t_n\right\rangle \left\langle t_n\middle|\omega_{m,\alpha}\right\rangle, \tag{4.24}$$

with the property:

$$\sum_{n=1}^{K} W(t_n, \omega_{m,\alpha}) = \frac{1}{K}\left|U(\omega_{m,\alpha})\right|^2 \sum_{n=1}^{K} \left|u(t_n)\right|^2, \tag{4.25}$$

$$\sum_{m=1}^{K} W(t_n, \omega_{m,\alpha}) = \frac{1}{K}\left|u(t_n)\right|^2 \sum_{m=1}^{K} \left|U(\omega_{m,\alpha})\right|^2. \tag{4.26}$$

**Example 6**: Approximate time derivative operator. As we show in paper II, in the continuous case, the time derivative is related to the angular frequency operstor via:

$$\frac{d}{dt}\langle t|u\rangle = \langle t|i\hat{\omega}|u\rangle. \tag{4.26}$$

We can use this relation as an approximate derivative at a point:

$$\begin{aligned} \frac{d}{dt}\langle t_k|u\rangle &\cong \langle t_k|i\hat{\omega}_\alpha|u\rangle = \sum_{n=1}^{K} \left\langle t\middle|\omega_{n,\beta}\right\rangle\left\langle\omega_{n,\beta}\right|i\hat{\omega}_\alpha|u\rangle \\ &= i\sum_{n=1}^{K}\sum_{m=1}^{K} \left\langle t\middle|\omega_{n,\beta}\right\rangle\left\langle\omega_{n,\beta}\middle|\omega_{m,\alpha}\right\rangle\omega_{m,\alpha}\left\langle\omega_{m,\alpha}\middle|u\right\rangle. \end{aligned} \tag{4.27}$$

There exists some arbitrariness in the choice of α and β, if they are equal, one gets:

$$\begin{aligned} \frac{d}{dt}\langle t_k|u\rangle &\cong \langle t_k|i\hat{\omega}_\alpha|u\rangle = i\sum_{n=1}^{K} \left\langle t\middle|\omega_{n,\alpha}\right\rangle\omega_{n,\alpha}\left\langle\omega_{n,\alpha}\middle|u\right\rangle \\ &= \frac{i}{\sqrt{K}}\sum_{n=1}^{K} \omega_{n,\alpha} e^{i\omega_{n,\alpha}t_k} U(\omega_{n,\alpha}). \end{aligned} \tag{4.28}$$

For real signals the result should be real. This is guaranteed for α=s, eq. (3.28).

## 5. Non orthonormal bases

In sec. 3 we assumed orthonormal bases for the Dirac space. They lead to simple inverse transformations, like eqs (3.18a) and (3.19). But that is not the only possibility, we may choose non orthonormal bases. In practice we may encounter situations where the bases will not be orthonormal from the beginning, like in the case of non uniform sampling and for spectral analysis with different windowing. Therefore we should discuss the non orthonormal bases case. We shall consider cases where the angular frequency base is non orthogonal but the time base orthogonal, frequency base orthogonal and time base non orthogonal, and when both bases are non orthogonal.

**Non orthogonal angular frequency base**. Let us start with an orthonormal time basis like in eqs. (3.4) and (3.6), and non-orthonormal angular frequency basis, for which the matrix W:

$$W_{kn} = \langle \omega_k | \omega_n \rangle, \quad k,n = 1,2,...K, \tag{5.1}$$

is different from the unit matrix I. The ket-vectors which form the basis have to be linearly independent There exist two criteria which can be used in order to check if vectors are linearly independent. In the first, we demand that the determinant, which columns are the vectors, is different from zero. In the second, the determinant of W (the Gram determinant) given by eq. (5.1), should be different from zero. Let us describe the first case. Let us define the matrix Ω who's columns are the basis ket-vectors:

$$\Omega_{kn} = \langle t_k | \omega_n \rangle, \tag{5.2}$$

where the index k denotes the k-th row. Let us also define the matrix $\Omega^{\dagger} \equiv \Omega^{h.c.}$ , the hermitian conjugate of $\Omega$, who's rows are the basis bra-vectors:

$$\Omega^{h.c.}_{kn} = \langle \omega_k | t_n \rangle, \tag{5.3}$$

and n is the index of the n-th column. The condition for the basis to be independent is that one of the following determinants is not equal to zero:

$$|\Omega| \neq 0, \quad \text{or} \quad |\Omega^{h.c.}| \neq 0, \quad \text{or} \quad |W| \neq 0. \tag{5.4}$$

The identity operator (the unit matrix ) has the following representation:

$$I = \sum_{k=1}^{K} \sum_{n=1}^{K} W^{-1}_{kn} |\omega_k\rangle\langle\omega_n| = \sum_{k=1}^{K} |\omega_k\rangle\langle\omega^{-1}_k| = \sum_{k=1}^{K} |\omega^{-1}_k\rangle\langle\omega_k|, \tag{5.5}$$

where $W^{-1}$ is the inverse matrix of W and $\left|\omega_k^{-1}\right\rangle$ is the reciprocal basis ket, satisfying:

$$\left\langle\omega_k\middle|\omega_n^{-1}\right\rangle = \left\langle\omega_k^{-1}\middle|\omega_n\right\rangle = \delta_{kn}. \tag{5.6}$$

Eq. (5.6) can be also written as:

$$\sum_{m=1}^{K} \left\langle\omega_k^{-1}\middle|t_m\right\rangle\left\langle t_m\middle|\omega_n\right\rangle = \sum_{m=1}^{K} \left\langle\omega_k\middle|t_m\right\rangle\left\langle t_m\middle|\omega_n^{-1}\right\rangle = \delta_{kn}, \tag{5.7a}$$

or equivalently:

$$\Omega^{-1}\Omega = \Omega^{h.c.}\left(\Omega^{h.c.}\right)^{-1} = I, \tag{5.7b}$$

i.e. the reciprocal basis bras are the rows of the inverse of Ω and the kets are the columns of the inverse of the hermitian conjugate of Ω.

Let us find the spectral representation of a signal u(t) with the eqs. (3.3-6), (3.8-9) and (5.1-7b):

$$u(t_k) \equiv \left\langle t_k\middle|u\right\rangle = \sum_{n=1}^{K} \left\langle t_k\middle|\omega_n\right\rangle\left\langle\omega_n^{-1}\middle|u\right\rangle, \tag{5.8}$$

from which the spectral amplitude, which is defined now as $\left\langle\omega_n^{-1}\middle|u\right\rangle$, can be evaluated, using eqs. (5.7a) and (5.7b):

$$U(\omega_n) \equiv \left\langle\omega_n^{-1}\middle|u\right\rangle = \sum_{k=1}^{K} \left\langle\omega_n^{-1}\middle|t_k\right\rangle\left\langle t_k\middle|u\right\rangle = \sum_{k=1}^{K} \Omega_{nk}^{-1}u(t_k). \tag{5.9}$$

Thus the evaluation of the spectral amplitude requires the inversion of the matrix Ω, which can be problematic if K, the number of data, is very large. Let us note that the transformation (3.8), with the requirement (3.10), is a unitary transformation, for which $\Omega^{-1} = \Omega^{h.c.}$, in this case eq. (5.9) is equal to eq. (3.18) and no inversion is needed. Thus the orthogonality of the basis greatly simplifies the calculations. When large matrices are involved and changes are small, instead of inverting matrices, one can use the following procedure. Define:

$$\Omega = \Omega_0 - \Delta\Omega, \tag{5.10}$$

Where $\Omega_0$ is a unitary matrix satisfying:

$${\Omega_0}^{-1} = {\Omega_0}^{h.c.}. \tag{5.11}$$

The matrix $\Omega^{-1}$ can then be expanded:

$$\begin{aligned}\Omega^{-1} &= (\Omega_0 - \Delta\Omega)^{-1} = \left[\Omega_0\left(I - \Omega_0^{-1}\Delta\Omega\right)\right]^{-1} = \left(I - \Omega_0^{-1}\Delta\Omega\right)^{-1}\Omega_0^{-1} \\ &= \left(I - \Omega_0^{h.c.}\Delta\Omega\right)^{-1}\Omega_0^{h.c.} = \left[I + \Omega_0^{h.c.}\Delta\Omega + \left(\Omega_0^{h.c.}\Delta\Omega\right)^2 + \dots\right]\Omega_0^{h.c.} \\ &= \Omega_0^{h.c.} + \Omega_0^{h.c.}\Delta\Omega\ \Omega_0^{h.c.} + \Omega_0^{h.c.}\Delta\Omega\ \Omega_0^{h.c.}\Delta\Omega\ \Omega_0^{h.c.}\Delta\Omega\ \Omega_0^{h.c.} + \dots\ .\end{aligned} \tag{5.12}$$

The series (5.12) can be computed using the following successive approximations:

$$\left[\Omega^{-1}\right]^{(n+1)} = \Omega_0^{h.c.} + \left[\Omega^{-1}\right]^{(n)}\Delta\Omega\ \Omega_0^{h.c.},\ \ \left[\Omega^{-1}\right]^{(1)} = \Omega_0^{h.c.}. \tag{5.13}$$

For small corrections the first (n=1), or the second approximation (n=2) may be sufficient.

In the following we will try to show that orthogonality is also related to stability. Let us consider as an example the signal (3.3) in the basis (3.4-6), and the new basis described by:

$$\Omega_{nk} = \langle t_n | \omega_k \rangle = \frac{1}{\sqrt{K}}\exp(i\omega_k t_n), \tag{5.14}$$

$$\omega_n = c\cdot(n-1)\cdot\Delta\omega, \qquad \Delta\omega = 2\pi(K\Delta t)^{-1}, \tag{5.15}$$

i.e. we recover the orthogonal basis (3.16) when c=1 and we scale the angular frequency basis with the factor c. Let us check for which values of c the new basis is independent. For that purpose we need to evaluate the determinant

$$|\Omega| = \frac{1}{\left(\sqrt{K}\right)^K}\begin{vmatrix} 1 & 1 & \dots & 1 \\ \exp(i\omega_2 t_1) & \exp(i\omega_2 t_2) & \cdots & \exp(i\omega_2 t_K) \\ \vdots & \vdots & \ddots & \vdots \\ \exp(i\omega_K t_1) & \exp(i\omega_K t_2) & \cdots & \exp(i\omega_K t_K)\end{vmatrix}, \tag{5.16}$$

which can vanish if either two columns or two rows are dependent. This may happen if:

$$\begin{aligned}&\exp(i\cdot c\cdot\Delta\omega\cdot t_k) = \exp(i\cdot c\cdot\Delta\omega\cdot t_n), \\ &\exp(i\cdot\Delta t\cdot\omega_m) = \exp(i\cdot\Delta t\cdot\omega_j),\end{aligned} \tag{5.17}$$

for some values of k, n, m and j. The conditions (5.13) may occur if:

$$c\cdot\Delta t\cdot\Delta\omega(k-n) = 2\pi\cdot M, \tag{5.18}$$

for some integers k and n not larger than K, and for some integer M.

From eqs. (5.15) and (5.18) we finally obtain that the determinant (5.16) will vanish if:

$$c = \frac{MK}{|k-n|}, \; 0 \le k \le K-1, \; 0 \le n \le K-1; \; k \ne n, \; M = 1,2,3,\dots \; . \qquad (4.19)$$

For example if K=100, (5.16) will be zero for

$$c = \frac{100}{99}, \; \frac{100}{98},\dots,\frac{100}{50}, \; \frac{200}{99}, \; \frac{200}{98},\dots\frac{200}{66}, \; \frac{300}{99}, \; \frac{300}{98},\dots, \; \frac{300}{75}, \; \frac{400}{99}, \; \text{etc'}.$$

For K=50

$$c = \frac{50}{49}, \; \frac{50}{48},\dots,\frac{50}{25}, \; \frac{100}{49}, \; \frac{100}{48},\dots,\frac{100}{33}, \; \frac{150}{49}, \; \frac{150}{48},\dots, \; \frac{150}{37}, \; \frac{200}{49}, \; \text{etc'}.$$

In figs. 2a and 2b the absolute value of the determinant $|\Omega|$ is displayed as a function of c for K=50.

**Non orthogonal time basis.** Let us consider the case of non uniform sampling, i.e. sampling for which the time intervals are not equal. Let us assume that the signal u(t) is sampled at times $t_k$ , k=0,1,…,K and that the average time interval is Δt. We can construct an orthonormal angular frequency basis exactly as in (3.16) with the identity operator (matrix) (3.17). We can use this basis in order to find an interpolation formula for the basis $|t_k\rangle$ , k=1,2,…,K. Assuming the transition (3.8), with the normalization (3.14) we arrive at :

$$|t_k\rangle = I|t_k\rangle = \sum_{n=0}^{K-1} \langle \omega_n | t_k \rangle |\omega_n\rangle = \frac{1}{\sqrt{K}} \sum_{n=0}^{K-1} \left[\exp(-i\omega_n t_k)\right] |\omega_n\rangle . \qquad (5.20)$$

Next, we can use the same procedure which led to eqs. (5.2), (5.3), (5.5) and (5.6), namely to define

$$\Omega_{kn} = \langle t_k | \omega_n \rangle, \qquad (5.21)$$

to find the reciprocity basis, and the identity operator (matrix) in the form:

$$I = \sum_{k=1}^{K} |t_k\rangle\langle t_k^{-1}| = \sum_{k=1}^{K} |t_k^{-1}\rangle\langle t_k| . \qquad (5.22)$$

The spectral representation of the signal will be:

$$u(t_k) \equiv \langle t_k | u \rangle = \sum_{n=1}^{K} \langle t_k | \omega_n \rangle \langle \omega_n | u \rangle, \qquad (5.23)$$

from which the spectral amplitude, which is defined now as $\langle \omega_n | u \rangle$, using eqs. (5.21) and (5.22), can be evaluated:

$$U(\omega_n) \equiv \langle \omega_n | u \rangle = \sum_{k=1}^{K} \langle \omega_n | t_k^{-1} \rangle \langle t_k | u \rangle = \sum_{k=1}^{K} \Omega_{nk}^{-1} u(t_k) . \qquad (5.24)$$

## 6. A variant of the sampling theorem

We shall derive a sampling theorem for a finite sampling. We formulate the problem in the following way. Let the Dirac space be spanned by the bases (3.4) and (3.16). We shall require that our signals (3.5) will depend only on angular frequencies

$$0 \le \omega_k \le \omega_B, \qquad 0 < B < K, \qquad (6.1)$$

where B is one of the allowed indices. The immediate consequence of eq. (6.1) is that the signals are spanned by a B dimensional angular frequency basis. It means that by the requirement (6.1) the signals are found in a lower (than K) dimensional space. Within the framework of the Dirac theory this can be achieved by using projection operators. Let us define the projection operator:

$$P = \sum_{k=0}^{B} |\omega_k\rangle\langle\omega_k| , \qquad (6.2)$$

which acts in the following way:

$$P|\omega_m\rangle = \sum_{k=0}^{B} |\omega_k\rangle\langle\omega_k|\omega_m\rangle = \sum_{k=0}^{B} |\omega_k\rangle\delta_{km} = \begin{cases} |\omega_m\rangle, & \text{if } m \le B, \\ 0, & \text{if } m > B. \end{cases} \qquad (6.3)$$

Thus P made a projection into a B dimensional subspace, let us call it $S_B$. We need only a B dimensional basis in the subspace $S_B$. The kets of eq. (6.3) can form such a basis. But what happens now to the basis (3.4)? The requirement (6.1) means that now all ket-vectors were projected to a B dimensional subspace $S_B$ with the aid of the projection operator (6.2). The basis (3.4) has also to be projected there:

$$P|t_m\rangle = \sum_{k=0}^{B} |\omega_k\rangle\langle\omega_k|t_m\rangle = \frac{1}{\sqrt{K}} \sum_{k=0}^{B} [\exp(-i\omega_k t_m)] |\omega_k\rangle \equiv |\tau_m\rangle . \qquad (6.4)$$

The kets $|\tau_m\rangle$, m=1,…,K, also form a basis in $S_B$, as any vector $|v\rangle$ in the subspace $S_B$ can be uniquely expanded with it.:

$$|v\rangle \equiv P|v\rangle = PIP|v\rangle = P\sum_{m=1}^{K} |t_m\rangle\langle t_m|P|v\rangle = \sum_{m=1}^{K} |\tau_m\rangle\langle\tau_m|v\rangle. \qquad (6.5)$$

While the ket-vectors $|\omega_k\rangle$, $0 \le k \le B$, form a complete basis in $S_B$, the ket-vectors (6.4) form an overcomplete basis, which we also called a coherent states basis.
Let us specify more carefully the meaning of completeness in the subspace $S_B$:

$$P = PIP = P\sum_{k=0}^{K-1}|\omega_k\rangle\langle\omega_k|P = \sum_{k=0}^{B}|\omega_k\rangle\langle\omega_k|, \qquad (6.6)$$

thus the projection operator becomes the identity operator in the subspace.

Let us pose the following question: is it possible to compose out of the K ket-vectors $|t_k\rangle$ of eq.(3.4), a complete B dimensional basis for $S_B$? In some circumstances, as it will be shown, it is feasible. Let us assume that K is even and

$$B=K/2-1, \qquad (6.7)$$

then:

$$\begin{aligned}
|\tau_m\rangle = IP|t_m\rangle &= \sum_{n=1}^{K}\sum_{k=0}^{B} |t_n\rangle\langle t_n|\omega_k\rangle\langle\omega_k|t_m\rangle \\
&= \frac{1}{K}\sum_{n=1}^{K}\sum_{k=0}^{B} \left[\exp(i\omega_k(t_n - t_m))\right]|t_n\rangle \\
&= \frac{1}{K}\sum_{n=1}^{K} \frac{\exp(i\cdot\Delta\omega\cdot(B+1)\cdot(t_n - t_m)) - 1}{\exp(i\cdot\Delta\omega\cdot(t_n - t_m)) - 1}|t_n\rangle \\
&= \frac{1}{K}\sum_{n=1}^{K} \frac{\exp[2\pi i(K/2)(n-m)/K] - 1}{\exp(i\cdot\Delta\omega\cdot(t_n - t_m)) - 1}|t_n\rangle \\
&= \frac{1}{K}\sum_{n=1}^{K} \frac{\exp[\pi i(n-m)] - 1}{\exp[2\pi i(n-m)/K] - 1}|t_n\rangle
\end{aligned} \qquad (6.8)$$

multiplying both sides by P we obtain:

$$|\tau_m\rangle = \frac{1}{K}\sum_{n=1}^{K} \frac{\exp[\pi i(n-m)] - 1}{\exp[2\pi i(n-m)/K] - 1}|\tau_n\rangle, \qquad (6.9)$$

In eq. (6.9) the ket-vector with odd index on the left hand side depends only on even indices in the sum on the right hand side and vice versa. Thus the projections of the basis $|t_n\rangle$ into the subspace $S_B$ form an overcomplete basis there, which can be split into two complete bases:

$$\text{(I):}\quad |\tau_1\rangle, |\tau_3\rangle, \ldots, |\tau_{K-1}\rangle, \tag{6.10}$$

$$\text{(II):}\quad |\tau_2\rangle, |\tau_4\rangle, \ldots, |\tau_K\rangle. \tag{6.11}$$

The sampling theorem can be now stated in the following two ways:

$$\langle\tau_{2m}|v\rangle = \frac{-2}{K}\sum_{n=1}^{K/2}\frac{1}{\exp[2\pi i(2m-2n+1)/K]-1}\langle\tau_{2n-1}|v\rangle,, \tag{6.12}$$

$$\langle\tau_{2m-1}|v\rangle = \frac{-2}{K}\sum_{n=1}^{K/2}\frac{1}{\exp[2\pi i(2m-2n-1)/K]-1}\langle\tau_{2n}|v\rangle, \tag{6.13}$$

where in eqs. (6.12) and (6.13) $m=1,\ldots,K/2$ .

## 7. Aliasing

Let us now study the symmetries of eq. (3.18.a). Let us note that:

$$e^{-i2\pi(\alpha+n-1)k/K} = e^{-i2\pi(\pm K+\alpha+n-1)k/K} = \left(e^{-i2\pi(\pm K-\alpha-n+1)k/K}\right)^*, \tag{7.1}$$

from which we can deduce the following symmetries of eq. (3.18a) for real signals:

$$\begin{aligned} U(\omega_{n,\alpha}) &= U(\omega_{n\pm K,\alpha}) = U^*(\omega_{-n\pm K,-\alpha}), \\ |U(\omega_{n,\alpha})| &= |U(\omega_{n\pm K,\alpha})| = |U(\omega_{-n\pm K,-\alpha})|. \end{aligned} \tag{7.2}$$

These symmetries in terms of frequencies f are:

$$\begin{aligned} U(f) &= U(f\pm f_S) = U(f\pm nf_S) = U^*(-f\pm f_S) = U^*(-f\pm mf_S), \\ |U(f)| &= |U(f\pm f_S)| = |U(-f\pm f_S)| = |U(\pm f\pm mf_S)|, \end{aligned} \tag{7.3}$$

where n and m are integers. The symmetries of eq. (7.3) are the cause of the well known phenomenon called aliasing, which occurs when a signal is undersampled [21-24]. The ket basis generated by the sampled times is insufficient to reproduce the whole signal and has artificial symmetries which are imposed on the spectral amplitude.

## 8. Some identities

As the number of orthonormal bases in angular frequencies is infinite one can get identities in the following way:

$$\begin{aligned}\langle\omega|u\rangle &= \langle\omega|\left(\sum_{n=1}^{K}|\omega_{n,\alpha}\rangle\langle\omega_{n,\alpha}|\right)|u\rangle = \sum_{n=1}^{K}\langle\omega|\omega_{n,\alpha}\rangle\langle\omega_{n,\alpha}|u\rangle \\ &= \sum_{k=1}^{K}\sum_{n=1}^{K}\langle\omega|t_k\rangle\langle t_k|\omega_{n,\alpha}\rangle\langle\omega_{n,\alpha}|u\rangle \\ &= \frac{1}{K}\sum_{n=1}^{K}\sum_{k=1}^{K}\left[\exp\left(it_k\left(\omega_{n,\alpha}-\omega\right)\right)\right]\langle\omega_{n,\alpha}|u\rangle \\ &= \frac{1}{K}\sum_{n=1}^{K}\exp\left(it_1\left(\omega_{n,\alpha}-\omega\right)\right)\frac{\exp\left(it_K\left(\omega_{n,\alpha}-\omega\right)\right)-1}{\exp\left(it_1\left(\omega_{n,\alpha}-\omega\right)\right)-1}\langle\omega_{n,\alpha}|u\rangle.\end{aligned} \tag{8.1}$$

They are the consequence of the expansion:

$$\begin{aligned}\langle\omega| &= \langle\omega|\left(\sum_{n=1}^{K}|\omega_{n,\alpha}\rangle\langle\omega_{n,\alpha}|\right) = \sum_{k=1}^{K}\sum_{n=1}^{K}\langle\omega|t_k\rangle\langle t_k|\omega_{n,\alpha}\rangle\langle\omega_{n,\alpha}| \\ &= \sum_{n=1}^{K}\exp\left(it_1\left(\omega_{n,\alpha}-\omega\right)\right)\frac{\exp\left(it_K\left(\omega_{n,\alpha}-\omega\right)\right)-1}{\exp\left(it_1\left(\omega_{n,\alpha}-\omega\right)\right)-1}\langle\omega_{n,\alpha}|.\end{aligned} \tag{8.2}$$

## 9. The uncertainty principle

A crud evaluation of the uncertainty of angular frequency determination can be obtained from eq. (3.14):

$$(K\Delta t)\Delta\omega=2\pi, \tag{9.1}$$

where $K\Delta t$ is the time window opening and $\Delta\omega$ is the grid length of the angular frequencies forming an orthogonal basis. If we treat the signal in a way similar to a wave function of quantum mechanics, we can obtain in a similar way the time frequency uncertainty principle. Let us consider the obvious inequality:

$$\begin{aligned}&\sum_{k=1}^{K}\left|\beta\left(t_k-\langle t\rangle\right)\nu u\left(t_k\right)+\left(\omega_{k,\alpha}-\langle\omega\rangle\right)U\left(\omega_{k,\alpha}\right)\right|^2 \\ &\quad=\beta^2\left\langle\Delta t^2\right\rangle-\beta B+\left\langle\Delta\omega^2\right\rangle\geq 0,\end{aligned} \tag{9.2}$$

where $\beta$ is a real auxiliary parameter and:

$$\langle t\rangle=\sum_{k=1}^{K} t_k\left|\nu u(t_k)\right|^2, \qquad \left\langle \Delta t^2\right\rangle=\sum_{k=1}^{K}\left(t_k-\langle t\rangle\right)^2\left|\nu u(t_k)\right|^2,$$
$$\langle \omega\rangle=\sum_{k=1}^{K} \omega_{k,\alpha}\left|U(\omega_{k,\alpha})\right|^2, \qquad \left\langle \Delta \omega^2\right\rangle=\sum_{k=1}^{K}\left(\omega_{k,\alpha}-\langle \omega\rangle\right)^2\left|U(\omega_{k,\alpha})\right|^2, \tag{9.3}$$
$$B=-2\sum_{k=1}^{K} \mathrm{Re}\left[\left(t_k-\langle t\rangle\right)\left(\omega_{k,\alpha}-\langle\omega\rangle\right)\left(\nu u(t_k)U(\omega_{k,\alpha})\right)\right].$$

One can consider the equality sign in eq. (9.2) and the resulting quadratic equation in $\beta$. The condition for an absence of real solutions in $\beta$ is that:

$$\left\langle \Delta t^2\right\rangle\left\langle \Delta \omega^2\right\rangle \geq B^2/4, \tag{9.4}$$

which is the uncertainty condition. More general discussions of uncertainty relations one can find in refs. [8] and [9].

## 10. Summary and conclusions

We have started section 2 with the basics of Dirac's representation theory and have shown its essentials by representing the bras and kets with rows and columns respectively. We have introduced there the notion of coherent states and discuss the consequences of restricting the vector space to a subspace. In sec. 3 ideas used in quantum mechanics, related to the properties of self-adjoint operators, especially the orthogonality of states, were employed in signal analysis. The physical states of quantum mechanics, corresponding to different eigenvalues, which are eigenstates of self-adjoint operators, are orthogonal. Therefore orthogonality means more than a convenient tool for expansion of functions.

In sec. 3 we introduced the "Dirac space" of dimension equal to the number of signal data. We gave an extensive analysis of all orthogonal bases labeled with time and angular frequencies respectively. The analogy to the quantum wave functions was fully employed and averaging quantities were proposed. In sec. 4, in a way similar to one used in quantum mechanics operators, observables and averages (expectation values) are defined. There we also introduced the signal and spectral operators respectively, and discussed their relation to the spectral amplitude and to double distributions. In sec. 5 we have discussed nonorthogonal bases and have noticed the peculiar stability of the orthogonal one. In sec. 6, we gave a new variant of the sampling theorem. We have decomposed the Dirac space into two subspaces of equal dimensions. The sampling theorem was the relation between the complete basis in a subspace and the overcomplete basis obtained by projecting the basis of the Dirac space into the subspace. In sec. 7 we derived and discussed the symmetries which were leading to the aliasing phenomenon. In sec. 8 some identities were derived, which were linking different angular frequency bases. In sec. 9 the uncertainty principle was derived..

In this paper we have demonstrated that the Dirac representation theory can be effectively adjusted and applied to signal theory. The main emphasis was put on orthogonality as the principal physical requirement. The particular role of the identity and projection operators was also stressed.

## Figure captions

1. The spectral power $P(\omega)$ dependence on frequency for the signal of eq. (3.24).

2a. The absolute value of the determinant $|\Omega|$ as a function of the scaling parameter c for K=50.

2b. Same as Fig. 2a, but with a semi logarithmic axes.

A. Gersten, Figure 1

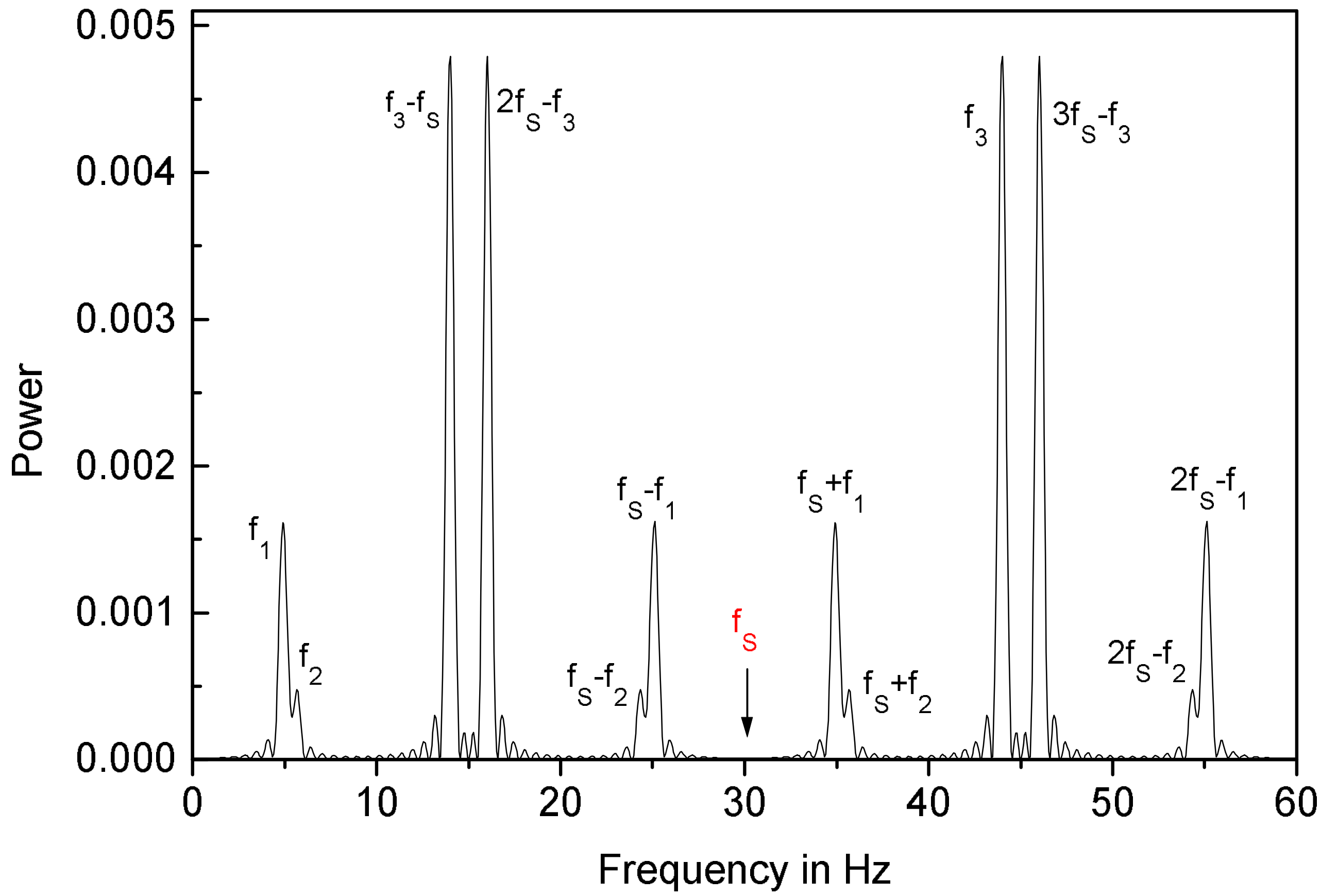

A. Gersten, Figure 2a

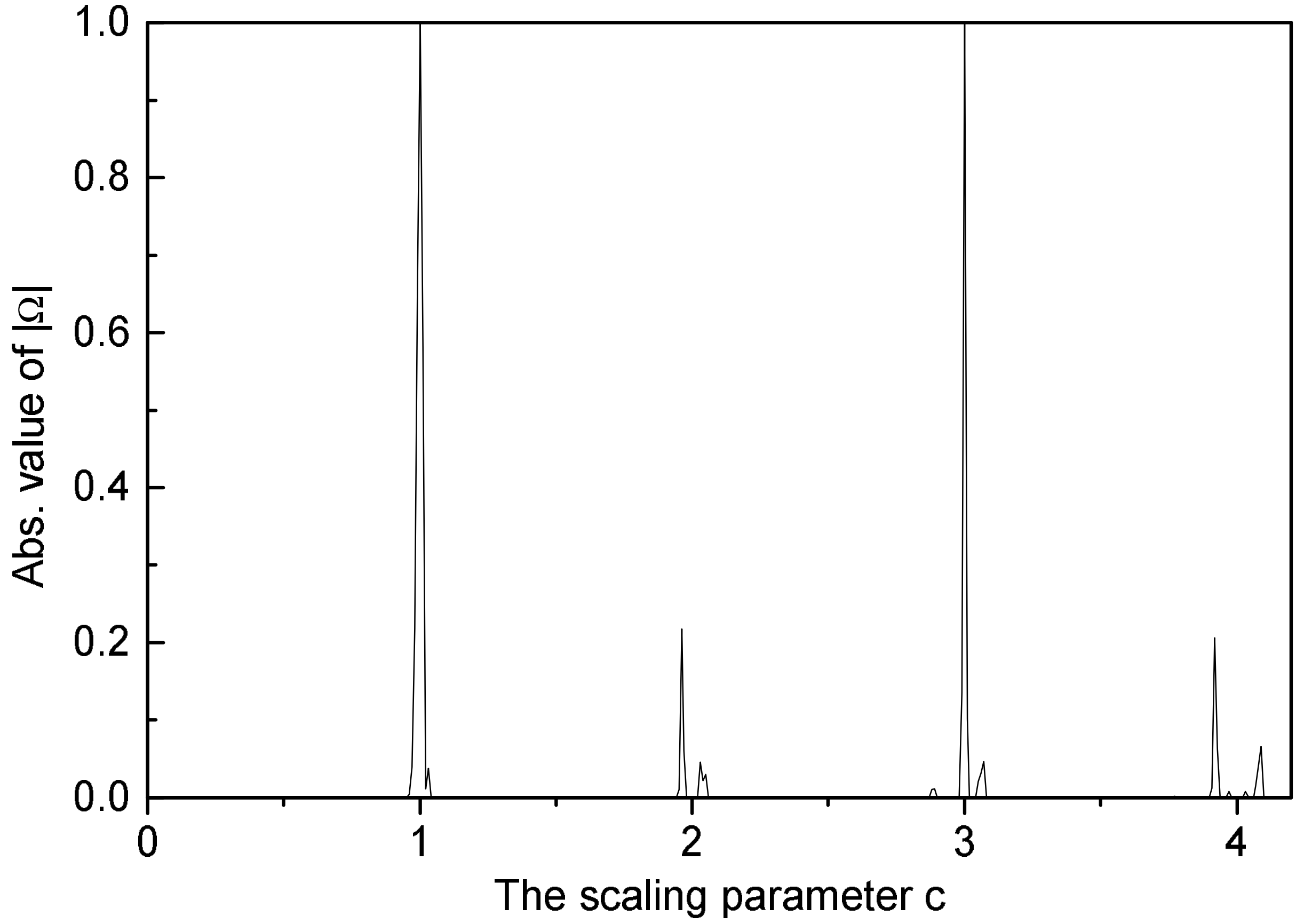

A. Gersten, Figure 2b

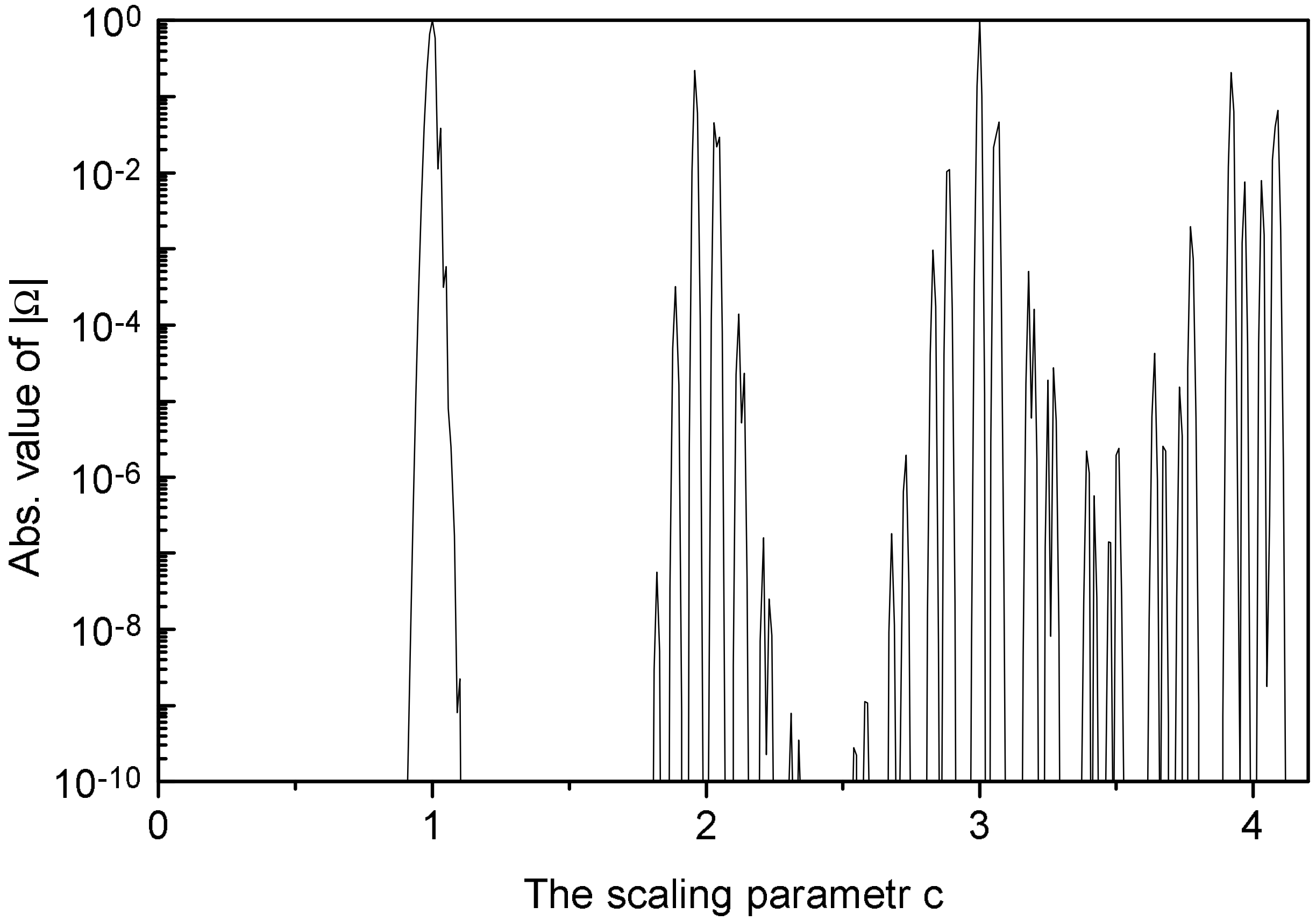